\documentclass{andp2012}
\usepackage[english]{babel}
\usepackage{graphicx}
\usepackage{dcolumn}

\usepackage{bbm}
\usepackage{mathrsfs}
\usepackage[bbgreekl]{mathbbol}
\usepackage{url}
\usepackage{hyperref}

\usepackage{subfigure}
\usepackage{multirow}

\usepackage{slashed}

\usepackage{esvect}

\renewcommand{\vec}[1]{\vv{#1}}

\newcommand{\nua}[1]{\ensuremath{\rlap
           {\kern-2.5pt\ensuremath
           {\overset{\scriptscriptstyle(-)}{\phantom{\nu}}}}
           {\ensuremath{{\nu}_{#1}}}}}

\newcommand{\fermion}{\ensuremath{\mathfrak{f}}}

\newcommand{\nff}{\ensuremath{f}}

\newcommand{\chg}{\ensuremath{e}}
\newcommand{\mgm}{\ensuremath{\mu}}
\newcommand{\elm}{\ensuremath{\epsilon}}
\newcommand{\anm}{\ensuremath{a}}

\newcommand{\elechg}{\ensuremath{e}}

\newcommand{\afl}{\ensuremath{\ell}}

\newcommand{\bmag}{\ensuremath{\mu_{\text{B}}}}

\keywords{Neutrino, Beyond Standard Model, Neutrino electromagnetic properties.}
\title{Electromagnetic neutrinos in laboratory experiments and astrophysics}
\author[C. Giunti]{Carlo Giunti\inst{1}}
\author[K. A. Kouzakov]{Konstantin A. Kouzakov\inst{3}}
\author[Y. Li]{Yu-Feng Li\inst{2}}
\author[A. V. Lokhov]{Alexey V. Lokhov\inst{6}}
\author[A. I. Studenikin]{Alexander I. Studenikin\inst{4,5,}\footnote{Corresponding author\quad E-mail:~\textsf{studenik@srd.sinp.msu.ru}}}
\author[Sh. Zhou]{Shun Zhou\inst{2}}
\address[1]{INFN, Sezione di Torino, and Dipartimento di Fisica Teorica, Universit`a di Torino}
\address[2]{Institite of High Energy Physics, Chinese Academy of Sciences, Beijing, China}
\address[3]{Department of Nuclear Physics and Quantum Theory of Collisions, Faculty of Physics, Lomonosov Moscow State University, 119991 Moscow, Russia}
\address[4]{Department of Theoretical Physics, Faculty of Physics, Lomonosov Moscow State University, 119991 Moscow, Russia}
\address[5]{Joint Institute for Nuclear Research, 141980 Dubna, Moscow Region, Russia}
\address[6]{Institute for Nuclear Research, Russian Academy of Sciences, 117312 Moscow, Russia}
\shortauthors{C. Giunti et al.}
\begin{abstract}
 An overview of neutrino electromagnetic properties, which open a door to
the new physics beyond the Standard Model, is given. The effects
of neutrino electromagnetic interactions both in terrestrial
experiments and in astrophysical environments are discussed. The
experimental bounds on neutrino electromagnetic characteristics
are summarized. Future astrophysical probes of electromagnetic
neutrinos are outlined.
\end{abstract} \shortabstract
\begin{document}
\maketitle

%
\section{Introduction}
\label{A001}
The importance of neutrino electromagnetic properties was first
mentioned by Pauli in 1930, when he postulated the existence of
this particle and discussed the possibility that the neutrino
might have a magnetic moment \cite{Pauli:1992mu}. Systematic
theoretical studies of neutrino electromagnetic properties started
after it was shown that in the extended Standard Model with
right-handed neutrinos the magnetic moment of a massive neutrino
is, in general, nonvanishing and that its value is determined by
the neutrino mass
\cite{Marciano:1977wx,Lee:1977tib,Fujikawa:1980yx,Petcov:1976ff,Pal:1981rm,Shrock:1982sc,Bilenky:1987ty}.

Neutrinos remained elusive until the detection of reactor
neutrinos by Reines and Cowan around 1956 \cite{Reines:1960pr}.
However, there was no sign of a neutrino mass. After the discovery
of parity violation in 1957, the
two-component theory of massless neutrinos was
proposed \cite{Landau:1957tp,Lee:1957qr,Salam:1957st}, in which a neutrino is
described by a Weyl spinor and there are only left-handed
neutrinos and right-handed antineutrinos. It was however clear
\cite{Touschek:1957,Case:1957zza,Mclennan:1957} that two-component
neutrinos could be massive Majorana fermions and that the
two-component theory of a massless neutrino is equivalent to the
Majorana theory in the limit of zero neutrino mass.

The two-component theory of massless neutrinos was later
incorporated in the Standard Model of Glashow, Weinberg and Salam
\cite{Glashow:1961tr,Weinberg:1967tq,Salam:1968rm}, in which
neutrinos are massless and have only weak interactions. In the
Standard Model Majorana neutrino masses are forbidden by the
$\text{SU}(2)_{L} \times \text{U}(1)_{Y}$ symmetry. We now know
that neutrinos are massive, because many experiments observed
neutrino oscillations (see the review articles
\cite{Giunti:2007ry,Bilenky:2010zza,Xing:2011zza,GonzalezGarcia:2012sz,Bellini:2013wra,Beringer:1900zz}),
which are generated by neutrino masses and mixing
\cite{Pontecorvo:1957cp,Pontecorvo:1957qd,Maki:1962mu,Pontecorvo1968}.
Therefore, the Standard Model must be extended to account for the
neutrino masses. There are many possible extensions of the
Standard Model which predict different properties for neutrinos
(see \cite{Ramond:1999vh,Mohapatra:2004,Xing:2011zza}). Among
them, most important is their fundamental Dirac or Majorana
character. In many extensions of the Standard Model neutrinos
acquire also electromagnetic properties through quantum loops'
effects which allow interactions of neutrinos with
electromagnetic fields and electromagnetic interactions of
neutrinos with charged particles.

Hence, the theoretical and experimental study of neutrino
electromagnetic interactions is a powerful tool in the search for
a more fundamental theory beyond the Standard Model. Moreover, the
electromagnetic interactions of neutrinos can generate important
effects, especially in astrophysical environments, where neutrinos
propagate over long distances in magnetic fields in vacuum and in
matter. Unfortunately, in spite of many efforts in the search of
neutrino electromagnetic interactions, up to now there is no
positive experimental indication in favor of their existence.
However, it is expected that the Standard Model neutrino charge
radii should be measured in the near future. This will be a test
of the Standard Model and of the physics beyond the Standard Model
which contributes to the neutrino charge radii. Moreover, the
existence of neutrino masses and mixing implies that neutrinos
have (diagonal and/or transition) magnetic moments. Since their values depend on the specific
theory which extends the Standard Model in order to accommodate
neutrino masses and mixing, experimentalists and theorists are
eagerly looking for them.

The paper is organized as follows. Section~\ref{C001} delivers the
general form of the electromagnetic interactions of Dirac and
Majorana neutrinos in the one-photon approximation, which are
expressed in terms of electromagnetic form factors. In
Section~\ref{S003} we discuss some basic processes which are
induced by the neutrino electromagnetic properties and some
important effects due to the interaction of neutrinos with
classical electromagnetic fields. In Section~\ref{S004} we
overview the experimental constraints on the neutrino electric and
magnetic moments, the electric charge (millicharge), the charge
radius and the anapole moment. In Section~\ref{S005} future
astrophysical probes of neutrino electromagnetic properties and
interactions are outlined. Finally, Section~\ref{H001} summarizes
this work.

\section{Neutrino electromagnetic characteristics}
\label{C001}
In this Section we discuss the general form of the electromagnetic
interactions of Dirac and Majorana neutrinos in the one-photon
approximation. In the Standard Model, the interaction of a
fermionic field ${\fermion}$ with the electromagnetic field
$A^{\mu}$ is given by the interaction Hamiltonian
\begin{equation}
\mathcal{H}_{\text{em}}^{(\fermion)} =
{j}_{\mu}^{(\fermion)}{A}^{\mu} = \chg_{\fermion}
\overline{\fermion} \gamma_{\mu} \fermion A^{\mu} ,
\label{C006}
\end{equation}
where $\chg_{\fermion}$ is the charge of the fermion $\fermion$.

\begin{figure}
\begin{center}
\includegraphics*[bb=235 681 360 767, width=0.5\linewidth]{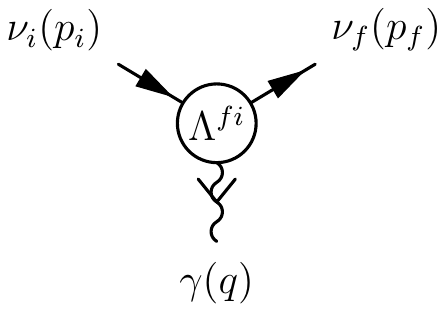}
\end{center}
\caption{ \label{C032} Effective one-photon coupling of neutrinos
with the electromagnetic field, taking into account possible
transitions between two different initial and final massive
neutrinos $\nu_{i}$ and $\nu_{f}$. }
\end{figure}

For neutrinos the electric charge is zero and there are no
electromagnetic interactions at tree-level\footnote{ However, in
some theories beyond the Standard Model neutrinos can be
millicharged particles (see below). }. At the same time, such interactions
can arise at the quantum level from loop diagrams at higher order
of the perturbative expansion of the interaction. We know that
there are at least three massive neutrino fields in nature, which are
mixed with the three active flavor neutrinos $\nu_{e}$,
$\nu_{\mu}$, $\nu_{\tau}$.
Therefore, we discuss the case of three
massive neutrino fields $\nu_{i}$ with respective masses
$m_{i}$ ($i=1,2,3$). In the one-photon approximation, the
effective electromagnetic interaction Hamiltonian is given by
\begin{equation}
\mathcal{H}_{\text{em}}^{(\nu)} = j_{\mu}^{(\nu)}A^{\mu}
= \sum_{i,f=1}^{3} \overline{\nu}_{f} \Lambda^{fi}_{\mu}
\nu_{i} A^{\mu} , \label{C033}
\end{equation}
where we take into account possible transitions between different
massive neutrinos. The physical effect of
$\mathcal{H}_{\text{em}}^{(\nu)}$ is described by the effective
electromagnetic vertex in Fig.~\ref{C032}. In momentum-space representation, this vertex depends
only on the four-momentum $q=p_i-p_f$ transferred to the photon
and can be expressed as follows:
\begin{align}
\Lambda_{\mu}(q) = \null & \null \left( \gamma_{\mu} - q_{\mu}
\slashed{q}/q^{2} \right) \left[ \nff_{Q}(q^{2}) + \nff_{A}(q^{2})
q^{2} \gamma_{5} \right] \nonumber
\\
\null & \null - i \sigma_{\mu\nu} q^{\nu} \left[ \nff_{M}(q^{2}) +
i \nff_{E}(q^{2}) \gamma_{5} \right] , \label{C043}
\end{align}
in which $\Lambda_{\mu}(q)$ is a $3{\times}3$ matrix in the space
of massive neutrinos expressed in terms of the four Hermitian
$3{\times}3$ matrices of form factors
\begin{equation}
\nff_{Q} = \nff_{Q}^{\dagger}, \qquad \nff_{M} = \nff_{M}^{\dagger},
\qquad \nff_{E} = \nff_{E}^{\dagger}, \qquad \nff_{A} = \nff_{A}^{\dagger},
\label{C044}
\end{equation}
where $Q,M,E,A$ refer respectively to the real charge, magnetic, electric, and anapole neutrino form factors. The Lorentz-invariant form of the vertex function~(\ref{C043}) is also consistent with electromagnetic gauge invariance that implies four-current conservation.

For the coupling with a real photon in vacuum ($q^{2}=0$) we have
\begin{equation}
\nff_{Q}^{fi}(0) = \chg_{fi} , \ \nff_{M}^{fi}(0) = \mgm_{fi} , \
\nff_{E}^{fi}(0) = \elm_{fi} , \ \nff_{A}^{fi}(0) = \anm_{fi} ,
\label{C045}
\end{equation}
where $\chg_{fi}$, $\mgm_{fi}$, $\elm_{fi}$ and $\anm_{fi}$ are,
respectively, the neutrino charge, magnetic moment, electric
moment and anapole moment of diagonal ($f=i$) and transition
($f{\neq}i$) types.

A Majorana neutrino is a neutral spin 1/2 particle which coincides
with its antiparticle. The four degrees of freedom of a Dirac
field (two helicities and two particle-antiparticle) are reduced
to two
(two helicities). 
Since a Majorana field has half the degrees of freedom of a Dirac
field, its electromagnetic properties are also reduced. Namely, in
the Majorana case the charge, magnetic and electric form-factor
matrices are antisymmetric and the anapole form-factor matrix is
symmetric. Since $\nff^{\text{M}}_{Q}$, $\nff^{\text{M}}_{M}$ and
$\nff^{\text{M}}_{E}$ are antisymmetric, a Majorana neutrino does
not have diagonal charge and dipole magnetic and electric form
factors \cite{Touschek:1957,Case:1957zza,Kayser:1982br}. It can only have a
diagonal anapole form factor. On the other hand, Majorana
neutrinos can have as many off-diagonal (transition) form-factors
as Dirac neutrinos.

{\bf Neutrino electric charge.} It is usually believed that the
neutrino electric charge $\chg_\nu=\nff_Q(0)$ is zero. This is
often thought to be attributed to the gauge-invariance and
anomaly-cancellation constraints imposed in the Standard Model. In
the Standard Model of SU(2)$_L \times $U(1)$_Y$ electroweak
interactions it is possible to get \cite{Foot:1992ui} a general
proof that neutrinos are electrically neutral, which is based on
the requirement of electric charges' quantization. The direct
calculations of the neutrino charge in the Standard Model for
massless (see, for instance
\cite{Bardeen:1972vi,CabralRosetti:1999ad}) and massive neutrinos
\cite{Dvornikov:2003js,Dvornikov:2004sj} also prove that, at
least at the one-loop level, the neutrino electric charge is
gauge-independent and vanishes. However, if the neutrino has a
mass, it still may become electrically millicharged. A brief
discussion of different mechanisms for introducing millicharged
particles including neutrinos can be found in
\cite{Davidson:2000hf}. In the case of millicharged massive neutrinos, electromagnetic gauge invariance implies that the diagonal electric charges $e_{ii}$ ($i=1,2,3$) are equal\footnote{The work is in preparation.}. It should be mentioned that the most
stringent experimental constraints on the electric charge of the
neutrino can be obtained from neutrality of matter. These and
other constraints, including the astrophysical ones, are discussed
in Section~\ref{S004}.

{\bf Neutrino charge radius.} Even if the electric charge of a
neutrino is zero, the electric form factor $\nff_Q(q^2)$ can still
contain nontrivial information about neutrino static properties
\cite{Giunti:2014ixa}. A neutral particle can be characterized by
a superposition of two charge distributions of opposite signs, so
that the particle form factor $\nff_Q(q^2)$ can be non-zero for
$q^2\neq 0$. The mean charge radius (in fact, it is the charged
radius squared) of an electrically neutral neutrino is given by
\begin{equation}\label{nu_cha_rad}
{\langle
r_{\nu}^2\rangle}={6}\left.\frac{d\nff_{Q}(q^2)}{dq^2}\right|_{
q^2=0},
\end{equation}
which is determined by the second term in the power-series
expansion of the neutrino charge form factor.

Note that there is a long-standing discussion (see
\cite{Giunti:2014ixa} for details) on the possibility to obtain
 for the neutrino charged radius a gauge-independent and
finite quantity. In one of the first studies \cite{Bardeen:1972vi}, it was
claimed that in the Standard Model and in the unitary gauge the
neutrino charge radius is ultraviolet-divergent and so it is not a
physical quantity.
However, it was shown
\cite{Lee:1973fw} that in the unitary gauge it is possible to obtain for the neutrino charge radius
a gauge-dependent but finite quantity.
Later on, it was also shown
\cite{Lee:1977tib} that considering additional box
diagrams in combination with contributions from the
proper diagrams it is possible to obtain a finite and
gauge-independent value for the neutrino charge radius.
In this
way, the neutrino electroweak radius was defined in
\cite{Lucio:1983mg,Lucio:1984jn} and an additional set of diagrams
that give contribution to its value was discussed in
\cite{Degrassi:1989ip}. Finally, in a series of papers
\cite{Bernabeu:2000hf,Bernabeu:2002nw,Bernabeu:2002pd} the
neutrino electroweak radius as a physical observable has been
introduced. This, however, revived the discussion
\cite{Fujikawa:2003tz,Fujikawa:2003ww,Papavassiliou:2003rx,Bernabeu:2004jr}
on the definition of the neutrino charge radius. Nevertheless, in the corresponding calculations, performed in
the one-loop approximation including additional terms from the
$\gamma-Z$ boson mixing and the box diagrams involving $W$ and $Z$
bosons, the following gauge-invariant result for the neutrino
charge radius has been obtained \cite{Bernabeu:2004jr}: ${\langle
r_{\nu_e}^2\rangle}=4 \times 10^{-33}$\,cm$^2$.
 This theoretical
result differs at most by an order of magnitude from the available
experimental bounds on $\langle r_{\nu}^2\rangle$ (see
Section~\ref{S004} for references and more detailed discussion).
Therefore, one may expect that the experimental accuracy will soon
reach the level needed to probe the neutrino effective charge
radius.

{\bf Neutrino electric and magnetic moments.} The most well
studied and understood among the neutrino electromagnetic
characteristics are the dipole magnetic and electric moments,
which are given by the corresponding form factors at $q^2=0$:
\begin{equation}
\mgm_{\nu}=\nff_{M}(0), \qquad \elm_{\nu}=\nff_{E}(0).
\end{equation}
The diagonal magnetic and electric moments of a Dirac neutrino in
the minimally-extended Standard Model with right-handed neutrinos,
derived for the first time in \cite{Fujikawa:1980yx}, are
respectively
\begin{equation}\label{mu_D}
    \mgm^{D}_{ii}
  = \frac{3e_0 G_F m_{i}}{8\sqrt {2} \pi ^2}\approx 3.2\times 10^{-19}
  \bmag\left(\frac{m_i}{1 \, \text{eV}}\right), \qquad \elm^{D}_{ii}=0,
  \end{equation}
where $\bmag$ is the Bohr magneton. According to (\ref{mu_D}) the
value of the neutrino magnetic moment is very small. However, in
many other theoretical frameworks (beyond the minimally-extended
Standard Model) the neutrino magnetic moment can reach values that
are of interest for the next generation of terrestrial experiments
and also accessible for astrophysical observations. Note that the
best laboratory upper limit on a neutrino magnetic moment,
$\mgm_{\nu} \leq 2.9 \times 10^{-11} \bmag$ (90\% CL), has been obtained by
the GEMMA collaboration \cite{Beda:2012zz} (see Section~\ref{S004}), and the best
astrophysical limit is $\mgm_{\nu}\leq 3 \times 10^{-12} \mgm _B$ (90\% CL)
\cite{Raffelt:1990pj}. The latter bound comes from the constraints on the possible delay of helium ignition of a red giant star in globular clusters due to the cooling induced by the energy loss in the plasmon-decay process $\gamma^*\to\nu\bar{\nu}$ (see Fig.~\ref{E003}).
Recently the limit has been updated in \cite{Viaux:2013hca}
using state-of-the-art astronomical observations and stellar evolution codes,
with the results
\begin{equation}
\mgm_{\nu}
<
\left\{
\begin{array}{l} \displaystyle
2.6 \times 10^{-12} \bmag
\quad
\text{(68\% CL)}
,
\\ \displaystyle
4.5 \times 10^{-12} \bmag
\quad
\text{(95\% CL)}
.
\end{array}
\right.
\label{E048}
\end{equation}
This astrophysical bound on a neutrino
magnetic moment is applicable to both Dirac and Majorana neutrinos
and constrains all diagonal and transition
dipole moments.

{\bf Neutrino anapole moment.} The notion of an anapole moment for
a Dirac particle was introduced by Zeldovich \cite{Zeldovich:1957zl} after
the discovery of parity violation. In order to understand the
physical characteristics of the anapole moment, it is useful to
consider its effect in the interactions with external
electromagnetic fields. The neutrino anapole moment contributes to
the scattering of neutrinos with charged particles. In order to
discuss its effects, it is convenient to consider strictly neutral
neutrinos with $\nff_{Q}(0)=0$ and define a reduced charge form
factor $\tilde{\nff}_{Q}(q^{2})$ such that
\begin{equation}
\nff_{Q}(q^{2}) = q^2 \, \tilde{\nff}_{Q}(q^{2}) . \label{G069}
\end{equation}
Then, from Eq.~(\ref{nu_cha_rad}), apart from a factor $1/6$, the
reduced charge form factor at $q^2=0$ is just the squared neutrino
charge radius:
\begin{equation}
\tilde{\nff}_{Q}(0) = \langle{r}^{2}_\nu\rangle / 6 . \label{G070}
\end{equation}
Let us now consider the charge and anapole parts of the neutrino
electromagnetic vertex function, 
as
\begin{equation}
\Lambda_{\mu}^{Q,A}(q) = \left( \gamma_{\mu} q^{2} - q_{\mu}
\slashed{q} \right) \left[ \tilde{\nff}_{Q}(q^{2}) +
\nff_{A}(q^{2}) \gamma_{5} \right] . \label{G071}
\end{equation}
Since for ultrarelativistic neutrinos the effect of $\gamma_{5}$
is only a
sign which depends on the helicity of the neutrino, 
the phenomenology of neutrino anapole moments is similar to that
of neutrino charge radii. Hence, the limits on the neutrino charge
radii discussed in Section~\ref{S004} apply also to the neutrino
anapole moments multiplied by a factor of 6.

\section{Basic electromagnetic processes of neutrinos}
\label{S003}
{\bf Neutrino-electron elastic scattering.} The most sensitive and
widely used method for the experimental investigation of the
neutrino magnetic moment is provided by direct laboratory
measurements of low-energy elastic scattering of neutrinos and
antineutrinos with electrons in reactor, accelerator and solar
experiments\footnote{ The effects of a neutrino magnetic moment in
other processes which can be observed in laboratory experiments
have been discussed in
\cite{Kim:1974xx,Kim:1978xk,Dicus:1978rz,Rosado:1982fr}. }.
Detailed descriptions of several experiments can be found in
\cite{Wong:2005pa,Beda:2007hf}.

Extensive experimental studies of the neutrino magnetic moment,
performed during many years, are stimulated by the hope to observe
a value much larger than the prediction in Eq.~(\ref{mu_D}) of the
minimally extended Standard Model with right-handed neutrinos. It
would be a clear indication of new physics beyond the extended
Standard Model. For example, the effective magnetic moment in
$\bar\nu_{e}$-$e$ elastic scattering in a class of extra-dimension
models can be as large as about $10^{-10} \bmag$
\cite{Mohapatra:2004ce}. Future higher precision reactor
experiments can therefore be used to provide new constraints on
large extra-dimensions.

The possibility for neutrino-electron elastic scattering due to
neutrino magnetic moment was first considered in
\cite{Carlson:1932rk} and the cross section of this process was
calculated in \cite{Bethe:1935cp} (for related short historical
notes see \cite{Kyuldjiev:1984kz}). Here we would like to recall
the paper by Domogatsky and Nadezhin \cite{Domogatsky:1971tu}, where the cross section of
\cite{Bethe:1935cp} was corrected and the antineutrino-electron
cross section was considered in the context of the earlier
experiments with reactor antineutrinos of
\cite{Cowan:1954pq,Cowan:1957pp}, which were aimed to reveal the
effects of the neutrino magnetic moment. Discussions on the
derivation of the cross section and on the optimal conditions for
bounding the neutrino magnetic moment, as well as a collection of
cross section formulae for elastic scattering of neutrinos
(antineutrinos) on electrons, nucleons, and nuclei can be found in
\cite{Kyuldjiev:1984kz,Vogel:1989iv}.

Let us consider the process
\begin{equation}
\nu_{\afl} + e^{-} \to \nu_{\afl'} + e^{-}, \label{D035}
\end{equation}
where a neutrino or antineutrino with flavor $\afl=e,\mu,\tau$ and energy $E_{\nu}$ elastically scatters off a free electron (FE) at rest in the laboratory frame. Due to neutrino mixing, the final neutrino flavor $\afl'$ can be different from $\afl$.
There are two observables: the kinetic energy $T_{e}$ of the
recoil electron and the recoil angle $\chi$ with respect to the
neutrino beam, which are related by
\begin{equation}\label{D036}
\cos\chi =
\frac{E_{\nu}+m_{e}}{E_{\nu}}\Big[\frac{T_{e}}{T_{e}+2m_{e}}\Big]^{1/2}
.
\end{equation}
The electron kinetic energy is constrained from the
energy-momentum conservation by
\begin{equation}
T_{e} \leq \frac {2E_{\nu}^{2}}{2E_{\nu} + m_{e}} .
\end{equation}

Since, in the ultrarelativistic limit, the neutrino magnetic
moment interaction changes the neutrino helicity and the Standard
Model weak interaction conserves the neutrino helicity, the two
contributions add incoherently in the cross section\footnote{ The
small interference term due to neutrino masses has been derived in
\cite{Grimus:1997aa}. } which can be written as
\cite{Vogel:1989iv},
\begin{equation}\label{D037}
\frac{d\sigma_{\nu_{\afl}e^{-}}}{dT_{e}} =
\left(\frac{d\sigma_{\nu_{\afl}e^{-}}}{dT_{e}}\right)_{\text{SM}}^{\text{FE}}
+
\left(\frac{d\sigma_{\nu_{\afl}e^{-}}}{dT_{e}}\right)_{\text{mag}}^{\text{FE}}
.
\end{equation}

The weak-interaction cross section is given by
\begin{align}
\left(\frac{d\sigma_{\nu_{\afl}e^{-}}}{dT_{e}}\right)_{\text{SM}}^{\text{FE}}
= \null & \null \frac{G^{2}_F m_{e}}{2\pi} \bigg\{
(g_{V}^{\nu_{\afl}} + g_{A}^{\nu_{\afl}})^{2} \nonumber
\\
\null & \null + (g_{V}^{\nu_{\afl}} - g_{A}^{\nu_{\afl}})^{2}
\left(1-\frac{T_{e}}{E_{\nu}}\right)^{2} \nonumber
\\
\null & \null + \left[ (g_{A}^{\nu_{\afl}})^{2} -
(g_{V}^{\nu_{\afl}})^{2} \right] \frac{m_{e}T_{e}}{E_{\nu}^{2}}
\bigg\} , \label{D038}
\end{align}
with the standard coupling constants $g_{V}$ and $g_{A}$ given by
\begin{align}
\null & \null g_{V}^{\nu_{e}} = 2\sin^{2} \theta_{W} + 1/2 , \quad
\null && \null g_{A}^{\nu_{e}} = 1/2 , \label{D039}
\\
\null & \null g_{V}^{\nu_{\mu,\tau}} = 2\sin^{2} \theta_{W} - 1/2
, \quad \null && \null g_{A}^{\nu_{\mu,\tau}} = - 1/2 .
\label{D040}
\end{align}
For antineutrinos one must substitute $g_A \to -g_A$.

The neutrino magnetic-moment contribution to the cross section is
given by \cite{Vogel:1989iv}
\begin{equation}
\left(\frac{d\sigma_{\nu_{\afl}e^{-}}}{dT_{e}}\right)_{\text{mag}}^{\text{FE}}
= \frac{\pi\alpha^{2}}{m_{e}^{2}}
\left(\frac{1}{T_{e}}-\frac{1}{E_{\nu}}\right)
\left(\frac{\mgm_{\nu_{\afl}}}{\bmag}\right)^{2} , \label{D041}
\end{equation}
where $\mgm_{\nu_{\afl}}$ is the effective magnetic moment
discussed in the following Section. It is called
traditionally ``magnetic moment'', but it receives
contributions from both the electric and magnetic dipole moments (see details in Section~\ref{S004}).

The two terms $(d\sigma_{\nu_{\afl}e^{-}}/dT_{e})_{\text{SM}}^{\text{FE}}$ and
$(d\sigma_{\nu_{\afl}e^{-}}/dT_{e})_{\text{mag}}^{\text{FE}}$ exhibit quite
different dependencies on the experimentally observable electron
kinetic energy $T_{e}$.
One can see that small values of the neutrino magnetic moment can
be probed by lowering the electron recoil energy threshold. In
fact, considering $T_{e} \ll E_{\nu}$ in Eq.~(\ref{D041}) and
neglecting the coefficients due to $g_{V}^{\nu_{\afl}}$ and
$g_{A}^{\nu_{\afl}}$ in Eq.~(\ref{D038}), one can find that
$(d\sigma/dT_{e})_{\text{mag}}^{\text{FE}}$ exceeds
$(d\sigma/dT_{e})_{\text{SM}}^{\text{FE}}$ for
\begin{equation}
T_{e} \lesssim \frac{\pi^{2}\alpha^{2}}{G_{\text{F}}^{2}m_{e}^3}
\left(\frac {\mgm_{\nu}}{\bmag}\right)^{2} .
\end{equation}

The current experiments with reactor antineutrinos have reached
threshold values of $T_e$ as low as few keV. As discussed in Section~\ref{H001}, these experiments are likely to
further improve the sensitivity to low energy deposition in the
detector. At low energies however one can expect a modification of
the free-electron formulas~(\ref{D038}) and~(\ref{D041}) due
to the binding of electrons in the germanium atoms, where e.g. the
energy of the $K_\alpha$ line, 9.89\,keV, indicates that at least
some of the atomic binding energies are comparable to the already
relevant to the experiment values of $T_e$. It was
demonstrated~\cite{Kouzakov:2010tx,Kouzakov:2011ig,Kouzakov:2011ka,Kouzakov:2011vx,Kouzakov:2011uq}
by means of analytical and numerical calculations that the atomic
binding effects are adequately described by the so-called stepping
approximation introduced in~\cite{kopeikin97} from interpretation
of numerical data. According to
the stepping approach, 
\begin{eqnarray}
\left(\frac{d\sigma_{\nu_{\afl}e^{-}}}{dT_{e}}\right)_{\text{SM}}=\left(\frac{d\sigma_{\nu_{\afl}e^{-}}}{dT_{e}}\right)_{\text{SM}}^{\text{FE}}\sum_j n_j\theta(T_e-E_j), \\
\left(\frac{d\sigma_{\nu_{\afl}e^{-}}}{dT_{e}}\right)_{\text{mag}}=\left(\frac{d\sigma_{\nu_{\afl}e^{-}}}{dT_{e}}\right)_{\text{mag}}^{\text{
FE}}\sum_j n_j\theta(T_e-E_j),
\end{eqnarray}
where the $j$ sum runs over all occupied atomic sublevels, with
$n_j$ and $E_j$ being their occupations and binding energies.

{\bf Neutrino-nucleus coherent scattering.} As mentioned above, the most
sensitive probe of neutrino electromagnetic properties is provided
by direct laboratory measurements of (anti)neutrino-electron
scattering at low energies in solar, accelerator and reactor
experiments (their detailed description can be found
in~\cite{Wong:2005pa, Balantekin:2006sw,
Beda:2007hf,Giunti:2008ve, Broggini:2012df,Giunti:2014ixa}). The coherent elastic neutrino-nucleus
scattering~\cite{Freedman:1973yd} has not been
experimentally observed so far, but it is expected to be
accessible in the reactor experiments when lowering the energy
threshold of the employed Ge
detectors~\cite{Wong:2011zzb,Li:2013fla,Li:2013ewa}.

Let us consider the case of electron neutrino scattering off a
spin-zero nucleus with even numbers of protons and neutrons, $Z$
and $N$. The matrix element of this process, taking into account
the neutrino electromagnetic properties, reads
\begin{eqnarray}
\label{M}
\mathcal{M}&=&\left[\frac{G_F}{\sqrt{2}}\bar{u}(k^\prime)\gamma^\mu(1-\gamma_5)u(k)C_V\right.\nonumber\\
&{}&+\frac{4\pi Ze}{q^2}\left({\chg_{\nu_e}}+\frac{e}{6}q^2\langle r_{\nu_e}^2\rangle\right)\bar{u}(k^\prime)\gamma^\mu u(k)\nonumber\\
&{}&\left.-\frac{4\pi
Ze\mgm_{\nu_e}}{q^2}\bar{u}(k^\prime)\sigma^{\mu\nu}q_\nu
u(k)\right]j_\mu^{(N)},
\end{eqnarray}
where $C_V=[Z(1-4\sin^2\theta_W)-N]/2$,
$j_\mu^{(N)}=(p_\mu+p_\mu^\prime)F(q^2)$, with $p$ and $p^\prime$
being the initial and final nuclear four-momenta. For neutrinos
with energies of a few MeV the maximum momentum transfer squared
($|q^2|_{\rm max}=4E_\nu^2$) is still small compared to $1/R^2$,
where $R$, the nucleus radius, is of the order of
$10^{-2}-10^{-1}$\,MeV$^{-1}$. Therefore, the nuclear elastic form
factor $F(q^2)$ can be set equal to one. Using (\ref{M}), one
obtains the differential cross section in the nuclear-recoil energy transfer
$T_N$ as a sum of two components. The first
component conserves the neutrino helicity and can be presented in
the form
\begin{equation}
\label{h.-c.}
\left(\frac{d\sigma_{\nu_{e}N}}{dT_{N}}\right)_{\text{SM}}^Q=\eta^2\,\left(\frac{d\sigma_{\nu_{e}N}}{dT_{N}}\right)_{\text{SM}},
\end{equation}
where
$$
\eta=1-\frac{\sqrt{2}\pi
eZ}{G_FC_V}\left[\frac{\chg_{\nu_e}}{MT}-\frac{e}{3}\langle
r_{\nu_e}^2\rangle\right],
$$
with $M$ being the nuclear mass, and
\begin{equation}
\label{SM}
\left(\frac{d\sigma_{\nu_{e}N}}{dT_{N}}\right)_{\text{SM}}=
\frac{G_F^2}{\pi}MC_V^2\left(1-\frac{T_N}{T_N^{\text{max}}}\right)
\end{equation}
is the Standard Model cross section due to weak
interaction~\cite{Drukier:1983gj}, with
$$
T_N^{\text{max}}=\frac{2E_\nu^2}{2E_\nu+M}.
$$
The second, helicity-flipping
component is due to the magnetic moment only and is given
by~\cite{Vogel:1989iv}
\begin{equation}
\label{NMM}
\left(\frac{d\sigma_{\nu_{e}N}}{dT_{N}}\right)_{\text{mag}}=4\pi
\alpha\mgm_{\nu_e}^2\,\frac{Z^2}{T_N}\left(1-\frac{T_N}{E_\nu}+\frac{T_N^2}{4E_\nu^2}\right).
\end{equation}

Clearly, any deviation of the measured cross section of the
process under discussion from the well-known Standard Model
value~(\ref{SM}) will provide a signature of the physics beyond
the Standard Model (see
also~\cite{Scholberg:2005qs,Barranco:2005yy,Barranco:2007tz,Davidson:2003ha}).
Formulas~(\ref{h.-c.}) and~(\ref{NMM}) describe such a deviation
due to neutrino electromagnetic interactions.

\begin{figure}
\null \hfill \subfigure[]{\label{E002}
\includegraphics*[bb=238 657 344 768, width=0.4\linewidth]{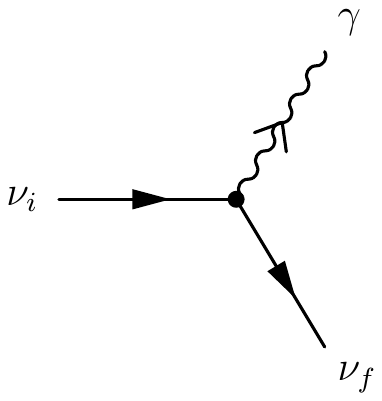}
} \hfill \subfigure[]{\label{E003}
\includegraphics*[bb=240 658 344 770, width=0.4\linewidth]{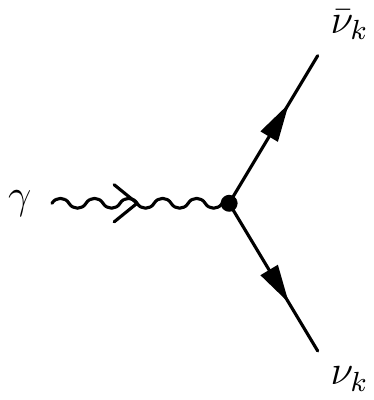}
} \hfill \null \caption{\label{E004} Feynman diagrams for neutrino
radiative decay and Cherenkov radiation \subref{E002} and plasmon
decay \subref{E003}. The depicted electromagnetic interaction vertices are supposed to be effective (such as the one-photon coupling in Fig.~\ref{C032}).}
\end{figure}

{\bf Radiative decay and related processes.} The magnetic and
electric (transition) dipole moments of neutrinos, as well as
possible very small electric charges (millicharges), describe
direct couplings of neutrinos with photons which induce several
observable decay processes. In this Section we discuss the decay
processes generated by the diagrams in Fig.~\ref{E004}: the
diagram in Fig.~\ref{E002} generates neutrino radiative decay
$\nu_{i}\to\nu_{f}+\gamma$ and the processes of neutrino Cherenkov
radiation and spin light ($SL\nu$) of a neutrino propagating in a
medium; the diagram in Fig.~\ref{E003} generates photon (plasmon)
decay to an neutrino-antineutrino pair in a plasma ($\gamma^{*}
\to \nu \bar\nu$).

If the masses of neutrinos are nondegenerate, the radiative decay
of a heavier neutrino $\nu_{i}$ into a lighter neutrino $\nu_{f}$
(with $m_{i}>m_{f}$) with emission of a photon,
\begin{equation}
\nu_{i} \to \nu_{f} + \gamma , \label{E006}
\end{equation}
may proceed in vacuum
\cite{Shrock:1974nd,Marciano:1977wx,Lee:1977tib,Petcov:1976ff,Goldman:1977jx,Zatsepin:1978iy,Pal:1981rm,Shrock:1982sc}.
Early discussions of the possible role of neutrino radiative decay
in different astrophysical and cosmological settings can be found
in
\cite{Dicus:1977nn,Sato:1977ye,Stecker:1980bu,Kimble:1980vz,Melott:1981iw,DeRujula:1980qd}.
The first estimates for the process of massive neutrino decay were
presented in \cite{Zatsepin:1978iy}. They considered various
processes of neutrino decay, for instance, the decay into three
neutrinos $\nu \rightarrow \nu + \nu + \bar{\nu}$ and the radiative
decay $\nu_1 \rightarrow \nu_2 + \gamma$.

In \cite{DeRujula:1980qd} the possible existence of relic slow
massive neutrinos was considered. The radiative decay of the
neutrino into an ultraviolet photon and a light neutrino becomes
then an indicator of these relic particles. The first one-loop
calculation of the neutrino radiative decay was performed in
\cite{Petcov:1976ff, Pal:1981rm} and yielded the decay rate as
\begin{equation}
\Gamma=\frac{\alpha
G_{F}^{2}}{128\pi^4}\left(\frac{m_1^2-m_2^2}{m_1}\right)^3(m_1^2+m_2^2)
\left|\sum_{\afl=e,\mu,\tau}U^{*}_{\afl 1}U_{\afl 2}r_\afl\right|^2,
\end{equation}
where $r_\afl \simeq 3({m_\afl }/2{m_W})^2$ ($\afl =e,\mu,\tau$), $m_W$
is the mass of W-boson, and $U_{\afl i}$ are the mixing-matrix elements.

The rate of neutrino radiative decay in relativistic
and non-relativistic media consisting of electrons (without muons
and taus)
was calculated in
Ref.~\cite{D'Olivo:1989un}
in the framework of finite-temperature quantum field theory.
The presence of the medium prevents the
Glashow-Iliopoulos-Maiani \cite{Glashow:1970gm} suppression of the decay,
which is strongly enhanced in high-density matter (neutron star,
supernova, etc.).
In Ref.~\cite{Grasso:1998td} the influence of dissipation and
dispersion in the medium,
that can be
important for the phenomenological studies of the early Universe,
was taken into account.
As shown in Ref.~\cite{Giunti:1990pp},
one can also calculate the rate of neutrino radiative decay in matter
avoiding the formalism of finite-temperature field theory
by considering
the effective averaged interaction with the medium.

{\bf Spin light of neutrino in matter.} The recent studies of
neutrino electromagnetic properties revealed a new mechanism of
electromagnetic radiation by a neutrino propagating in dense
matter that has been proposed in \cite{Lobanov:2002ur}. This type
of electromagnetic radiation was called the spin light of neutrino
in matter ($SL\nu$). In a quasi-classical treatment this radiation
is due to neutrino magnetic moment precession in dense background
matter. The quantum theory of this phenomena has been developed in
\cite{Studenikin:2004dx,Lobanov:2005zn,Grigoriev:2005bc}.

The $SL\nu$ is a process of photon emission in neutrino transition
between different helicity states in matter. As it has been shown
in Ref.~\cite{Studenikin:2004dx,Grigoriev:2005bc}, in the
relativistic regime the $SL\nu$ mechanism could provide up to one
half of the initial neutrino energy transition to the emitted
radiation. It was also shown that the $SL\nu$ provides the spin
polarization effect of neutrino beam moving in matter (similarly
to the well-known effect of the electron spin self-polarization in
synchrotron radiation \cite{Sokolov:1963zn}).

The characteristics of the $SL\nu$ depend on the components of the
medium. The $SL\nu$ is radiated by a neutrino with negative
helicity while propagating in matter consisting of electrons. In
the medium consisting of neutrons the $SL\nu$ is produced by an
antineutrino with positive helicity.

For the relativistic neutrinos the radiation is focused into a
narrow cone in the direction of the initial neutrino. The
radiation of ultra-relativistic neutrinos in matter has circular
polarization which in some cases (high density) reaches 100\%.
The average energy of the radiated photons depends on the energy
of the initial neutrinos and in dense medium reaches one half of
the initial neutrino energy (see also
\cite{Grigoriev:2011rq,Grigoriev:2012pw}).

Along with studying the conventional spin light of neutrino in
matter with the mass for the initial and final neutrino states one
can consider the spin light process in neutrino transition between
different mass states with masses $m_1$ and $m_2$, $m_1 > m_2$.
The emitted photon is coupled to the neutrinos by the transition
magnetic moment $\mu_{fi}$. To avoid cumbersome formulae, the
effects of oscillations were neglected and the matter with only
a neutron component was considered ($n_n \gg n_e \approx n_p$). It
was shown \cite{Grigoriev:2010ni,Grigoriev:2010uk} that the rate
of $SL\nu$ in the neutrino radiative decay acquires additional terms
that are proportional to the difference of the initial and final
neutrino masses squared: $\delta = \frac{m_1^2-m_2^2}{p_1^2}$. As
opposed to the $SL\nu$ with the same mass of the initial and final
state, the process is kinematically open for the quasi-vacuum case,
when the density of the background medium is small. In addition, the
expression for the rate of the process can be reduced to the
results of previous neutrino radiative decay calculations. The
influence of external fields and matter on a massive neutrino
decay was further considered in \cite{Ternov:2013ana}

{\bf Neutrino interaction with electromagnetic fields.} If
neutrinos have nontrivial electromagnetic properties, they can
interact with classical electromagnetic fields. Significant
effects can occur, in particular, in neutrino astrophysics, since
neutrinos can propagate over very long distances in astrophysical
environments with magnetic fields. In this case even a very weak
interaction can have large cumulative effects.

A classical electromagnetic field produces spin and spin-flavor
neutrino transitions
\cite{Cisneros:1971nq,Fujikawa:1980yx,Voloshin:1986ty,Okun:1986na,Akhmedov:1988nc,Lim:1988tk}.
This kind of interaction can yield observable effects, for
instance, in the solar neutrino data
\cite{Akhmedov:2002mf,Chauhan:2003wr,Miranda:2003yh,Miranda:2004nz,Balantekin:2004tk,Guzzo:2005rr,Friedland:2005xh,Yilmaz:2008vh}.
The neutrino effective magnetic moment is also modified in very
strong external magnetic fields \cite{Borisov:1985ha}. It has been
recently shown that due to the nontrivial electromagnetic
properties the production of neutrino-antineutrino pairs becomes
possible in very strong magnetic fields \cite{Gavrilov:2012aw}. It
can be also important to account for the external electromagnetic
fields and for the background matter simultaneously. In various
astrophysical situations the effects of fields and matter can
either cancel or enhance each other.

For instance, an approach based on the generalized
Bargmann-Michel-Telegdi equation can be used for derivation of an
impact of matter motion and polarization on the neutrino spin (and
spin-flavor) evolution. Consider, as an example, an electron
neutrino spin precession in the case when neutrinos with the
Standard Model interaction are propagating through moving and
polarized matter composed of electrons (electron gas) in the
presence of an electromagnetic field given by the
electromagnetic-field tensor $F_{\mu \nu}=({\vec E}, {\vec B})$.
As discussed in \cite{Studenikin:2004bu} (see also
\cite{Egorov:1999ah,Lobanov:2001ar}) the evolution of the
three-di\-men\-sio\-nal neutrino spin vector $\vec S $ is given by
\begin{equation}\label{Spin} {d\vec S
\over dt}={2\mgm\over \gamma} \Big[ {\vec S \times ({\vec
B_0}+\vec M_0)} \Big],
\end{equation}
where the magnetic field $\vec{B_0}$ in the neutrino rest frame is
determined by the transversal and longitudinal (with respect to
the neutrino motion) magnetic and electric field components in the
laboratory frame,
 \begin{equation}
\vec B_0=\gamma\Big(\vec B_{\perp} +{1 \over \gamma} \vec
B_{\parallel} + \sqrt{1-\gamma^{-2}} \Big[\vec E_{\perp} \times
\frac{\vec \beta}{\beta} \Big]\Big).
\end{equation}
The matter term $\vec M_0$ in Eq. (\ref{Spin}) is also composed of
the transversal $\vec {M}{_{0_{\parallel}}}$ and longitudinal
$\vec {M_{0_{\perp}}}$ parts,
\begin{equation}
\vec {M_0}=\vec {M}{_{0_{\parallel}}}+\vec {M_{0_{\perp}}},
\label{M_0}
\end{equation}
\begin{equation}
\begin{array}{c}
\displaystyle \vec {M}_{0_{\parallel}}=\gamma\vec\beta{n_{0} \over
\sqrt {1- v_{e}^{2}}}\left\{ \rho^{(1)}_{e}\left(1-{{\vec v}_e
\vec\beta \over {1- {\gamma^{-2}}}} \right)\right. \\-
\displaystyle\rho^{(2)}_{e}\left. \left(\vec\zeta_{e}\vec\beta
\sqrt{1-v^2_e}+ {(\vec \zeta_{e}{\vec v}_e)(\vec\beta{\vec v}_e)
\over 1+\sqrt{1-v^2_e} }\right){1 \over {1- {\gamma^{-2}}}}
\right\}, \label{M_0_parallel}
\end{array}
\end{equation}
\begin{equation}\label{M_0_perp}
\begin{array}{c}
\displaystyle \vec {M}_{0_{\perp}}=-\frac{n_{0}}{\sqrt {1-
v_{e}^{2}}}\Bigg\{ \vec{v}_{e_{\perp}}\Big(
\rho^{(1)}_{e}+\rho^{(2)}_{e}\frac {(\vec{\zeta}_{e}{\vec{v}_e})}
{1+\sqrt{1-v^2_e}}\Big) \\+ \displaystyle
{\vec{\zeta}_{e_{\perp}}}\rho^{(2)}_{e}\sqrt{1-v^2_e}\Bigg\}.
\end{array}
\end{equation}
Here $n_0=n_{e}\sqrt {1-v^{2}_{e}}$ is the invariant number
density of matter given in the reference frame for which the total
speed of matter is zero. The vectors $\vec v_e$, and $\vec \zeta_e
\ (0\leqslant |\vec \zeta_e |^2 \leqslant 1)$ denote,
respectively, the speed of the reference frame in which the mean
momentum of matter (electrons) is zero, and the mean value of the
polarization vector of the background electrons in the above
mentioned reference frame. The coefficients $\rho^{(1,2)}_e$ are
calculated if the neutrino Lagrangian is given, and  within the
extended Standard Model supplied with $SU(2)$-singlet right-handed
neutrino $\nu_{R}$,
\begin{equation}\label{rho}
\rho^{(1)}_e={\tilde{G}_F \over {2\sqrt{2}\mu }}\,, \qquad
\rho^{(2)}_e =-{G_F \over {2\sqrt{2}\mu}}\,,
\end{equation}
where $\tilde{G}_{F}={G}_{F}(1+4\sin^2 \theta _W).$ For the
probability of the neutrino spin oscillations in the adiabatic
approximation we get from Eqs. (\ref{M_0_parallel}) and
(\ref{M_0_perp})
\begin{equation}\label{ver2}
P_{\nu_L \rightarrow \nu_R} (x)=\sin^{2} 2\theta_\textmd{eff}
\sin^{2}{\pi x \over L_\textmd{eff}},
\end{equation}
\begin{equation}
\sin^{2} 2\theta_\textmd{eff}={E^2_\textmd{eff} \over
{E^{2}_\textmd{eff}+\Delta^{2}_\textmd{eff}}}, \ \ \
L_\textmd{eff}={2\pi \over
\sqrt{E^{2}_\textmd{eff}+\Delta^{2}_\textmd{eff}}},
\end{equation}
where
\begin{equation}\label{E3}
E_\textmd{eff}=\mgm \Big|{\vec B}_{\perp} + {1\over \gamma}{\vec
M}_{0\perp} \Big|,
\end{equation}
\begin{equation}\label{Delta3}
\Delta^{2}_\textmd{eff}={\mgm \over \gamma}\Big|{\vec
M}_{0\parallel}+{\vec B}_{0\parallel} \Big|.
\end{equation}
It follows that even without presence of an electromagnetic field,
${\vec B}_{\perp}={\vec B}_{0\parallel}=0$, neutrino spin (or
spin-flavor) oscillations can be induced in the presence of matter
when the transverse matter term ${\vec M}_{0\perp}$ is not zero.
This possibility is realized in the case of nonzero transversal
matter velocity or polarization. A detailed discussion of this
phenomenon can be found in \cite{Studenikin:2004bu,
Studenikin:2004tv}.

\section{Experimental limits on neutrino electromagnetic properties}
\label{S004}

{\bf Effective magnetic moment.} In scattering experiments the
neutrino is created at some distance from the detector as a flavor
neutrino, which is a superposition of massive neutrinos.
Therefore, the magnetic moment that is measured in these
experiments is not that of a single massive neutrino, but it is an
effective magnetic moment which takes into account neutrino mixing
and the oscillations during the propagation between source and
detector \cite{Grimus:1997aa,Beacom:1999wx}. In the following,
when we refer to an effective magnetic moment of a flavor neutrino
without indication of a source-detector distance $L$ it is
implicitly understood that $L$ is small, such that the effective
magnetic moment is independent of the neutrino energy and from the
source-detector distance. In such a case, the effective magnetic moment is given by~\cite{Giunti:2014ixa}
\begin{equation}
\mgm_{\nu_{\afl}}^{2}
\simeq
\mgm_{\bar\nu_{\afl}}^{2}
\simeq
\sum_{f=1}^3
\left|
\sum_{i=1}^3
U_{\afl i}^{*}
\left(
\mgm_{fi} - i \elm_{fi}
\right)
\right|^2.
\label{D050}
\end{equation}
%

Another situation where the effective
magnetic moment does not depend on the neutrino energy and on the
source-detector distance is when the source-detector distance is
much larger than all the oscillation lengths $L_{fi} = 4 \pi
E_{\nu} / |\Delta{m}^{2}_{fi}|$. The effective magnetic moment in this case is evaluated as~\cite{Giunti:2014ixa}
\begin{equation}
\mgm_{\nu_{\afl}}^{2}
\simeq
\mgm_{\bar\nu_{\afl}}^{2}
\simeq
\sum_{i=1}^3
|U_{\afl i}|^2
\sum_{f=1}^3
\left|
\mgm_{fi} - i \elm_{fi}
\right|^2.
\label{D053}
\end{equation}
Note that in the case of solar
neutrinos, which have been used by the Super-Kamiokande
\cite{Liu:2004ny} and Borexino \cite{Arpesella:2008mt} experiments
to search for neutrino magnetic moments, one must take into
account the matter effects. The latter can be done by replacing the neutrino mixing matrix in Eq.~(\ref{D053}) with the effective mixing matrix in matter at the point of neutrino production inside the Sun (see~\cite{Giunti:2014ixa} and references therein).

It is also interesting to note that
flavor neutrinos can have effective magnetic moments even if
massive neutrinos are Majorana particles. In this case, since massive Majorana
neutrinos do not have diagonal magnetic and electric dipole moments, the effective magnetic moments of
flavor neutrinos receive contributions only from the transition
dipole moments.

\begin{table*}
\renewcommand{\arraystretch}{1.2}
\begin{tabular}{lllll}
Method & Experiment & Limit & CL & Reference\\
\hline \multirow{5}{*}{Reactor $\bar\nu_e$-$e^-$}
&Krasnoyarsk        &$\mgm_{\nu_e} < 2.4 \times 10^{-10}\,\bmag$    &90\%   &\cite{Vidyakin:1992nf}     \\
&Rovno          &$\mgm_{\nu_e} < 1.9 \times 10^{-10}\,\bmag$    &95\%   &\cite{Derbin:1993wy}       \\
&MUNU           &$\mgm_{\nu_e} < 9   \times 10^{-11}\,\bmag$    &90\%   &\cite{Daraktchieva:2005kn} \\
&TEXONO         &$\mgm_{\nu_e} < 7.4 \times 10^{-11}\,\bmag$    &90\%   &\cite{Wong:2006nx}     \\
&GEMMA          &$\mgm_{\nu_e} < 2.9 \times 10^{-11}\,\bmag$    &90\%   &\cite{Beda:2012zz}     \\
\hline \multirow{1}{*}{Accelerator $\nu_e$-$e^-$}
&LAMPF          &$\mgm_{\nu_e} < 1.1 \times 10^{-9}\,\bmag$     &90\%   &\cite{Allen:1992qe}        \\
\hline \multirow{1}{*}{Accelerator
($\nu_{\mu},\bar\nu_{\mu}$)-$e^-$}
&BNL-E734       &$\mgm_{\nu_{\mu}} < 8.5 \times 10^{-10}\,\bmag$    &90\%   &\cite{Ahrens:1990fp}       \\
&LAMPF          &$\mgm_{\nu_{\mu}} < 7.4 \times 10^{-10}\,\bmag$    &90\%   &\cite{Allen:1992qe}        \\
&LSND           &$\mgm_{\nu_{\mu}} < 6.8 \times 10^{-10}\,\bmag$    &90\%   &\cite{Auerbach:2001wg}     \\
\hline \multirow{1}{*}{Accelerator
($\nu_{\tau},\bar\nu_{\tau}$)-$e^-$}
&DONUT          &$\mgm_{\nu_{\tau}} < 3.9 \times 10^{-7}\,\bmag$    &90\%   &\cite{Schwienhorst:2001sj} \\
\hline \multirow{2}{*}{Solar $\nu_e$-$e^-$}
&Super-Kamiokande   &$\mgm_{\text{S}}(E_{\nu} \gtrsim 5 \, \text{MeV}) < 1.1 \times 10^{-10}\,\bmag$    &90\%   &\cite{Liu:2004ny}      \\
&Borexino       &$\mgm_{\text{S}}(E_{\nu} \lesssim 1 \, \text{MeV}) < 5.4 \times 10^{-11}\,\bmag$   &90\%   &\cite{Arpesella:2008mt}    \\
\hline
\end{tabular}
\caption{\label{D055} Experimental limits for different neutrino
effective magnetic moments. }
\end{table*}

The constraints on the neutrino magnetic moments in direct
laboratory experiments have been obtained so far from the lack of
any observable distortion of the recoil electron energy spectrum.
Experiments of this type have started in the 50's at the Savannah
River Laboratory where the ${\bar\nu_{e}}$-$e^{-}$ elastic
scattering process was studied
\cite{Cowan:1954pq,Cowan:1957pp,Reines:1976pv} with somewhat
controversial results, as discussed by \cite{Vogel:1989iv}. The
most significant experimental limits on the effective magnetic
moment $\mgm_{\nu_e}$ which have been obtained in measurements of reactor
${\bar\nu_{e}}$-$e^{-}$ elastic scattering after about 1990 are listed in
Tab.~\ref{D055} (some details of the different experimental setups
are reviewed in \cite{Broggini:2012df}).

An attempt to improve the experimental bound on $\mgm_{\nu_e}$ in
reactor experiments was undertaken in \cite{Wong:2010pb}, where it
was suggested that in $\bar\nu_{e}$ interactions on an atomic
target the atomic electron binding (``atomic-ionization effect'')
can significantly increase the electromagnetic contribution to the
differential cross section with respect to the free-electron
approximation. However, the dipole approximation used to derive
the atomic-ionization effect is not valid for the electron
antineutrino cross section in reactor neutrino magnetic moment
experiments. Instead, the free electron approximation is
appropriate for the interpretation of the data of reactor neutrino
experiments and the current constraints in Tab.~\ref{D055} cannot
be improved by considering the atomic electron binding
\cite{Voloshin:2010vm,Kouzakov:2010tx,Kouzakov:2011ig,Kouzakov:2011ka,Kouzakov:2011vx,Kouzakov:2011uq,Chen:2013lba}.
The history and present status of the theory of neutrino-atom
collisions is reviewed in \cite{Kouzakov:2014lka}.

The current best limit on $\mgm_{\nu_e}$ has been obtained in 2012
in the GEMMA experiment at the Kalinin Nuclear Power Plant
(Russia) with a 1.5 kg highly pure germanium detector exposed at a
$\bar\nu_{e}$ flux of $2.7 \times 10^{13} \, \text{cm}^{-2} \,
\text{s}^{-1}$ at a distance of 13.9 m from the core of a $3 \,
\text{GW}_{\text{th}}$ commercial water-moderated reactor
\cite{Beda:2012zz}. The competitive TEXONO experiment is based at
the Kuo-Sheng Reactor Neutrino Laboratory (Taiwan), where a 1.06
kg highly pure germanium detector was exposed to the flux of
$\bar\nu_{e}$ at a distance of 28 m from the core of a $2.9 \,
\text{GW}_{\text{th}}$ commercial reactor
\cite{Wong:2006nx}\footnote{ The TEXONO and GEMMA data have been
also used by \cite{Barranco:2011wx,Healey:2013vka} to constrain
neutrino nonstandard interactions. }.

Searches for effects of neutrino magnetic moments have been
performed also in accelerator experiments. The LAMPF bounds on
$\mgm_{\nu_{e}}$ in Tab.~\ref{D055} have been obtained with
$\nu_{e}$ from $\mu^+$ decay \cite{Allen:1992qe}. The LAMPF and
LSND bounds on $\mgm_{\nu_{\mu}}$ in Tab.~\ref{D055} have been
obtained with $\nu_{\mu}$ and $\bar\nu_{\mu}$ from $\pi^+$ and
$\mu^+$ decay \cite{Allen:1992qe,Auerbach:2001wg}. The DONUT
collaboration \cite{Schwienhorst:2001sj} investigated
$\nu_{\tau}$-$e^-$ and $\bar\nu_{\tau}$-$e^-$ elastic scattering,
finding the limit on $\mgm_{\nu_{\tau}}$ in Tab.~\ref{D055}.

Solar neutrino experiments can also search for a neutrino magnetic
moment signal by studying the shape of the electron spectrum
\cite{Beacom:1999wx}.
Table~\ref{D055} gives the limits obtained in the Super-Kamiokande
experiment \cite{Liu:2004ny} for
\begin{eqnarray}
\mgm_{\text{S}}^{2}(E_{\nu} \gtrsim
5 \, \text{MeV})
&\simeq&
\cos^2\vartheta_{13}
\sum_{i=1}^3
\left|
\mgm_{i2} - i \elm_{i2}
\right|^2\nonumber\\
&{}&+
\sin^2\vartheta_{13}
\sum_{i=1}^3
\left|
\mgm_{i3} - i \elm_{i3}
\right|^2,
\label{D063}
\end{eqnarray}
where $\vartheta_{13}$ is the mixing angle,
and that obtained in the Borexino experiment
\cite{Arpesella:2008mt} for
\begin{equation}
\mgm_{\text{S}}(E_{\nu} \lesssim 1 \, \text{MeV})
\simeq
\mgm_{\nu_{e}},
\label{D062}
\end{equation}
where $\mgm_{\nu_{e}}$ is given by Eq.~(\ref{D053}).

Information on neutrino magnetic moments has been obtained also
with global fits of solar neutrino data
\cite{Joshipura:2002bp,Grimus:2002vb,Tortola:2004vh}. Considering
Majorana three-neutrino mixing, the authors of
\cite{Tortola:2004vh} obtained, at 90\% CL,
\begin{equation}
\sqrt{ |\mgm_{12}|^2 + |\mgm_{23}|^2 + |\mgm_{31}|^2 } < 4.0
\times 10^{-10} \, \bmag , \label{D065}
\end{equation}
from the analysis of solar and KamLAND, and
\begin{equation}
\sqrt{ |\mgm_{12}|^2 + |\mgm_{23}|^2 + |\mgm_{31}|^2 } < 1.8
\times 10^{-10} \, \bmag , \label{D066}
\end{equation}
adding the Rovno \cite{Derbin:1993wy}, TEXONO \cite{Li:2002pn} and
MUNU \cite{Daraktchieva:2003dr} constraints.

The neutrino magnetic moment contribution to the
(anti)neutrino-electron elastic scattering process flips the
neutrino helicity. If neutrinos are Dirac particles, this process
transforms active left-handed neutrinos into sterile right-handed
neutrinos, leading to dramatic effects on the explosion of a
core-collapse supernova
\cite{Dar:1987yv,Nussinov:1987zr,Goldman:1987fg,Lattimer:1988mf,Barbieri:1988nh,Notzold:1988kz,Voloshin:1988xu,Ayala:1998qz,Ayala:1999xn,Balantekin:2007xq},
where there are also contributions from the (anti)neutrino-proton and
(anti)neutrino-neutron elastic scattering. Requiring that the entire
energy in a supernova collapse is not carried away by the escaping
sterile right-handed neutrinos created in the supernova core, the
authors of \cite{Ayala:1998qz,Ayala:1999xn} obtained the following
upper limit on a generic neutrino magnetic moment:
\begin{equation}
\mgm_{\nu} \lesssim \left( 0.1 - 0.4 \right) \times 10^{-11} \,
\bmag , \label{D067}
\end{equation}
which is slightly more stringent than the bound $ \mgm_{\nu}
\lesssim \left( 0.2 - 0.8 \right) \times 10^{-11} \, \bmag $
obtained in \cite{Barbieri:1988nh}.

There is a gap of many orders of magnitude between the present
experimental limits on neutrino magnetic moments of the order of
$10^{-11} \, \bmag$ and the prediction smaller than about
$10^{-19} \, \bmag$ in Eq.~(\ref{mu_D}) of the minimal extension
of the Standard Model with right-handed neutrinos. The hope to
reach in the near future an experimental sensitivity of this order
of magnitude is very small, taking into account that the
experimental sensitivity of reactor ${\bar\nu_{e}}$-$e$ elastic
scattering experiments has improved by only one order of
magnitude during a period of about twenty years (see
\cite{Vogel:1989iv}, where a sensitivity of the order of
$10^{-10}\bmag$ is discussed). However, the experimental studies
of neutrino magnetic moments are stimulated by the hope that new
physics beyond the minimally extended Standard Model with
right-handed neutrinos might give much stronger contributions.

{\bf Neutrino millicharge.} The most severe experimental
constraint on neutrino electric charges is that on the effective
electron neutrino charge $\chg_{\nu_{e}}$, which can be obtained
from electric charge conservation in neutron beta decay $n \to p +
e^- + \bar\nu_{e}$, from the experimental limits on the
non-neutrality of matter which constrain the sum of the proton and
electron charges, $\chg_{p} + \chg_{e}$, and from the experimental
limits on the neutron charge $\chg_{n}$
\cite{Raffelt:1996wa,Raffelt:1999gv}. Several experiments which
measured the neutrality of matter give their results in terms of
\begin{equation}
\chg_{\text{mat}} = \frac{Z (\chg_{p} + \chg_{e}) + N \chg_{n}}{A}
, \label{G030}
\end{equation}
where $A=Z+N$ is the atomic mass of the substance under study, $Z$
is its atomic number and $N$ is its neutron number. From electric
charge conservation in neutron beta decay, we have
\begin{equation}
\chg_{\nu_{e}} = \chg_{n} - (\chg_{p} + \chg_{e}) = \frac{A}{Z}
\left( \chg_{n} - \chg_{\text{mat}} \right) . \label{G031}
\end{equation}
The best recent bound on the non-neutrality of matter
\cite{Bressi:2011pj},
\begin{equation}
\chg_{\text{mat}} = (- 0.1 \pm 1.1) \times 10^{-21} \, \elechg ,
\label{G032}
\end{equation}
has been obtained with $\text{S}\text{F}_{6}$, which has
$A=146.06$ and $Z=70$. Using the independent measurement of the
charge of the free neutron \cite{Baumann:1988ue}
\begin{equation}
\chg_{n} = (- 0.4 \pm 1.1) \times 10^{-21} \, \elechg ,
\label{G033}
\end{equation}
we obtain
\begin{equation}
\chg_{\nu_{e}} = (-0.6 \pm 3.2) \times 10^{-21} \, \elechg .
\end{equation}
This value is compatible with the neutrality of matter limit in
Tab.~\ref{G029}, which has been derived
\cite{Raffelt:1996wa,Raffelt:1999gv} from the value of $\chg_{n}$
in Eq.~(\ref{G033}) and $ \chg_{\text{mat}} = (0.8 \pm 0.8) \times
10^{-21} \, \elechg $ \cite{Marinelli:1983nd}.

It is also interesting that the effective charge of $\bar\nu_{e}$
can be constrained by the SN 1987A neutrino measurements taking
into account that galactic and extragalactic magnetic field can
lengthen the path of millicharged neutrinos and requiring that
neutrinos with different energies arrive on Earth within the
observed time interval of a few seconds \cite{Barbiellini:1987zz}:
\begin{equation}
|\chg_{\nu_{e}}| \lesssim 3.8 \times 10^{-12}
\frac{(E_{\nu}/10\,\text{MeV})}{(d/10\,\text{kpc})
(B/1\,\mu\text{G})} \sqrt{
\frac{\Delta{t}/t}{\Delta{E_{\nu}}/E_{\nu}} } , \label{G034}
\end{equation}
considering a magnetic field $B$ acting over a distance $d$ and
the corresponding time $t=d/c$. $E_{\nu} \approx 15\,\text{MeV}$
is the average neutrino energy, $\Delta{E_{\nu}} \approx E_{\nu} /
2$ is the energy spread, and $\Delta{t} \approx 5 \, \text{s}$ is
the arrival time interval. The authors of \cite{Barbiellini:1987zz} considered two
cases:

\begin{enumerate}

\item An intergalactic field $B \approx 10^{-3} \, \mu\text{G}$
acting over the whole path $d \simeq 50\,\text{kpc}$, which
corresponds to $t \simeq 5 \times 10^{12} \, \text{s}$, gives
\begin{equation}
|\chg_{\nu_{e}}| \lesssim 2 \times 10^{-15} \, \elechg .
\label{G035}
\end{equation}

\item A galactic field $B \approx 1 \, \mu\text{G}$ acting over a
distance $d \simeq 10\,\text{kpc}$, which corresponds to $t \simeq
1 \times 10^{12} \, \text{s}$, gives
\begin{equation}
|\chg_{\nu_{e}}| \lesssim 2 \times 10^{-17} \, \elechg .
\label{G036}
\end{equation}

\end{enumerate}

\begin{table*}
\renewcommand{\arraystretch}{1.2}
\begin{tabular}{lll}
Limit & Method & Reference\\
\hline
$|\chg_{\nu_{\tau}}| \lesssim 3 \times 10^{-4}\,\elechg$        &SLAC $e^{-}$ beam dump             &\cite{Davidson:1991si} \\
$|\chg_{\nu_{\tau}}| \lesssim 4 \times 10^{-4}\,\elechg$        &BEBC beam dump                 &\cite{Babu:1993yh}     \\
$|\chg_{\nu}| \lesssim 6 \times 10^{-14}\,\elechg$      &Solar cooling (plasmon decay)          &\cite{Raffelt:1999gv}  \\
$|\chg_{\nu}| \lesssim 2 \times 10^{-14}\,\elechg$      &Red giant cooling (plasmon decay)      &\cite{Raffelt:1999gv}  \\
$|\chg_{\nu_e}| \lesssim 3 \times 10^{-21}\,\elechg$    &Neutrality of matter               &\cite{Raffelt:1999gv}  \\
$|\chg_{\nu_e}| \lesssim 3.7 \times 10^{-12}\,\elechg$  &Nuclear reactor                &\cite{Gninenko:2006fi} \\
$|\chg_{\nu_e}| \lesssim 1.5 \times 10^{-12}\,\elechg$  &Nuclear reactor                &\cite{Studenikin:2013my}   \\
\hline
\end{tabular}
\caption{\label{G029} Approximate limits for different neutrino
effective charges. The limits on $\chg_{\nu}$ apply to all
flavors. }
\end{table*}

The last two limits in Tab.~\ref{G029} have been obtained
\cite{Gninenko:2006fi,Studenikin:2013my} considering the results
of reactor neutrino magnetic moment experiments. The differential
cross section of the $\bar\nu_{e}$--$e^{-}$ elastic scattering
process due to a neutrino effective charge $\chg_{\nu_{e}}$ is
given by (see \cite{Berestetskii:1979aa})
\begin{equation}
\left(\frac{d\sigma}{dT_{e}}\right)_{\text{charge}} \simeq
2\pi\alpha \frac{1}{m_{e}T_{e}^2}\chg_{\nu_{e}}^2 . \label{G037}
\end{equation}
In reactor experiments the neutrino magnetic moment is searched by
considering data with $T_{e} \ll E_{\nu}$. The ratio of
the charge cross section (\ref{G037}) and the magnetic moment
cross section in Eq.~(\ref{D041}), for which we consider only the
dominant part proportional to $1/T_{e}$, is given by
\begin{equation}
R = \frac{ \left(d\sigma/dT_{e}\right)_{\text{charge}} }{
\left(d\sigma/dT_{e}\right)_{\text{mag}} } \simeq
\frac{2m_e}{T_{e}} \frac{ \left( \chg_{\nu_{e}} / \elechg
\right)^2 }{ \left( \mgm_{\nu_{e}} / \bmag \right)^2 }
\label{G038}
\end{equation}
Considering an experiment which does not observe any effect of
$\mgm_{\nu_{e}}$ and obtains a limit on $\mgm_{\nu_{e}}$, it is
possible to obtain, following \cite{Studenikin:2013my}, a bound on
$\chg_{\nu_{e}}$ by demanding that the effect of $\chg_{\nu_{e}}$
is smaller than that of $\mgm_{\nu_{e}}$, i.e. that $R \lesssim 1$:
\begin{equation}
\chg_{\nu_{e}}^{2} \lesssim \frac{T_{e}}{2m_e}
\left(\frac{\mgm_{\nu_{e}}}{\bmag}\right)^{2} \elechg^2 .
\label{G039}
\end{equation}
The last limit in Tab.~\ref{G029} has been obtained from the 2012
results \cite{Beda:2012zz} of the GEMMA experiment, considering
$T_{e}$ at the experimental threshold of $2.8 \, \text{keV}$.

Let us finally note that a strong limit on a generic neutrino
electric charge $\chg_{\nu}$ can be obtained by considering the
influence of millicharged neutrinos on the rotation of a
magnetized star which is undergoing a core-collapse supernova
explosion (the neutrino star turning mechanism, $\nu$ST)
\cite{Studenikin:2012vi}. During the supernova explosion, the
escaping millicharged neutrinos move along curved orbits inside
the rotating magnetized star and slow down the rotation of the
star. This mechanism could prevent the generation of a rapidly
rotating pulsar in the supernova explosion. Imposing that the
frequency shift of a forming pulsar due to the neutrino star
turning mechanism is less than a typical observed frequency of
$0.1 \, \text{s}^{-1}$ and assuming a magnetic field of the order
of $10^{14} \, \text{G}$, the author of
\cite{Studenikin:2012vi} obtained
\begin{equation}\label{G040}
|\chg_{\nu}| \lesssim 1.3 \times 10^{-19} \, \elechg .
\end{equation}
Note that this limit is much stronger than the astrophysical
limits in Tab.~\ref{G029}.

\begin{table*}
\renewcommand{\arraystretch}{1.2}
\begin{tabular}{llcll}
Method & Experiment & Limit [$\text{cm}^2$] & CL & Reference\\
\hline \multirow{2}{*}{Reactor $\bar\nu_e$-$e^-$}
&Krasnoyarsk    &$|\langle{r_{\nu_e}^{2}}\rangle|<7.3\times10^{-32}$            &90\%   &\cite{Vidyakin:1992nf}     \\
&TEXONO     &$-4.2\times10^{-32}<\langle{r_{\nu_e}^{2}}\rangle<6.6\times10^{-32}$   &90\%   &\cite{Deniz:2009mu}    \\
\hline \multirow{2}{*}{Accelerator $\nu_e$-$e^-$} &LAMPF
&$-7.12\times10^{-32}<\langle{r_{\nu_e}^{2}}\rangle<10.88\times10^{-32}$
&90\%   &\cite{Allen:1992qe}\\
&LSND       &$-5.94\times10^{-32}<\langle{r_{\nu_e}^{2}}\rangle<8.28\times10^{-32}$     &90\%   &\cite{Auerbach:2001wg} \\
\hline \multirow{2}{*}{Accelerator $\nu_{\mu}$-$e^-$}
&BNL-E734   &$-4.22\times10^{-32}<\langle{r_{\nu_{\mu}}^{2}}\rangle<0.48\times10^{-32}$ &90\%   &\cite{Ahrens:1990fp}   \\
&CHARM-II   &$|\langle{r_{\nu_{\mu}}^{2}}\rangle|<1.2\times10^{-32}$                &90\%   &\cite{Vilain:1994hm}   \\
\hline
\end{tabular}
\caption{\label{G042} Experimental limits for the electron
neutrino charge radius. In the TEXONO, LAMPF, LSND, BNL-E734, and
CHARM-II cases, the published limits are half, because they use a
convention which differs by a factor of 2 (see also
Ref.~\cite{Hirsch:2002uv}).}
\end{table*}

{\bf Neutrino charge radius.} The neutrino charge radius has an
effect in the scattering of neutrinos with charged particles. The
most useful process is the elastic scattering with electrons.
Since in the ultrarelativistic limit the charge form factor
conserves the neutrino helicity, a neutrino charge radius
contributes to the weak interaction cross section
$(d\sigma/dT_{e})_{\text{SM}}$ of $\nu_{\afl}$--$e^{-}$ elastic
scattering through the following shift of the vector coupling
constant $g_{V}^{\nu_{\afl}}$
\cite{Grau:1985cn,Degrassi:1989ip,Vogel:1989iv,Hagiwara:1994pw}:
\begin{equation}
g_{V}^{\nu_{\afl}} \to g_{V}^{\nu_{\afl}} + \frac{2}{3} m_{W}^{2}
\langle{r}_{\nu_{\afl}}^{2}\rangle \sin^{2}\theta_{W} .
\label{G054}
\end{equation}
Using this method, experiments which measure neutrino-electron
elastic scattering can probe the neutrino charge radius. Some
experimental results are listed in Tab.~\ref{G042}. In addition,
the authors of \cite{Hirsch:2002uv} obtained the following 90\% CL bounds on
$\langle{r_{\nu_{\mu}}^{2}}\rangle$ from a reanalysis of CHARM-II
\cite{Vilain:1994hm} and CCFR \cite{McFarland:1997wx} data:
\begin{equation}
-0.52\times10^{-32}<\langle{r_{\nu_{\mu}}^{2}}\rangle<0.68\times10^{-32}
\, \text{cm}^2 . \label{G055}
\end{equation}
More recently, the authors of \cite{Barranco:2007ea} obtained the
following 90\% CL bounds on $\langle{r_{\nu_e}^{2}}\rangle$ from a
combined fit of all available $\nu_{e}$--$e^-$ and
$\bar\nu_{e}$--$e^-$ data:
\begin{equation}
-0.26\times10^{-32}<\langle{r_{\nu_e}^{2}}\rangle<6.64\times10^{-32}
\, \text{cm}^2. \label{G056}
\end{equation}

The single photon production process $e^{+} + e^{-} \to \nu +
\bar\nu + \gamma$ has been used to get bounds on the effective
$\nu_{\tau}$ charge radius, assuming a negligible contribution of
the $\nu_{e}$ and $\nu_{\mu}$ charge radii
\cite{Altherr:1993hb,Tanimoto:2000am,Hirsch:2002uv}. For Dirac
neutrinos, the authors of \cite{Hirsch:2002uv} obtained
\begin{equation}
-5.6\times10^{-32}<\langle{r_{\nu_{\tau}}^{2}}\rangle<6.2\times10^{-32}
\, \text{cm}^2 . \label{G057}
\end{equation}

Comparing the theoretical Standard Model values 
with the experimental limits in Tab.~\ref{G042} and those in
Eqs.~(\ref{G055})--(\ref{G057}), one can see that they differ at
most by one order of magnitude. Therefore, one may expect that the
experimental accuracy will soon reach the value needed to probe
the Standard Model predictions for the neutrino charge radii. This
will be an important test of the Standard Model calculation of the
neutrino charge radii. If the experimental value of a neutrino
charge radius
is found to be different from the Standard Model prediction 
it will be necessary to clarify the precision of the theoretical
calculation in order to understand if the difference is due to new
physics beyond the Standard Model.

The neutrino charge radius has also some impact on astrophysical
phenomena and on cosmology. The limits on the cooling of the Sun
and white dwarfs due to the plasmon decay process discussed in the
previous Section induced by a neutrino charge radius led the authors of
\cite{Dolgov:1981hv} to estimate the respective limits
$|\langle{r}_{\nu}^{2}\rangle| \lesssim 10^{-28} \, \text{cm}^{2}$
and $|\langle{r}_{\nu}^{2}\rangle| \lesssim 10^{-30} \,
\text{cm}^{2}$ for all neutrino flavors. From the cooling of red
giants the authors of \cite{Altherr:1993hb} inferred the limit
$|\langle{r}_{\nu}^{2}\rangle| \lesssim 4 \times 10^{-31} \,
\text{cm}^{2}$.

If neutrinos are Dirac particles, $e^{+}$--$e^{-}$ annihilations
can produce right-handed neutrino-antineutrino pairs through the
coupling induced by a neutrino charge radius. This process would
affect primordial Big-Bang Nucleosynthesis and the energy release
of a core-collapse supernova. From the measured $^4\text{He}$
yield in primordial Big-Bang Nucleosynthesis the authors of
\cite{Grifols:1986ed} obtained
\begin{equation}
|\langle{r}_{\nu}^{2}\rangle| \lesssim 7 \times 10^{-33} \,
\text{cm}^{2} , \label{G058}
\end{equation}
and from SN 1987A data the authors of \cite{Grifols:1989vi}
obtained
\begin{equation}
\langle{r}_{\nu}^{2}\rangle \lesssim 2 \times 10^{-33} \,
\text{cm}^{2} , \label{G059}
\end{equation}
for all neutrino flavors.

\section{Future astrophysical probes of electromagnetic neutrinos}
\label{S005}
{\bf Solar neutrinos.} The precision measurements of low-energy
neutrinos from the Sun in the ongoing and forthcoming solar
neutrino experiments will not only provide us with more accurate
values of neutrino oscillation parameters~\cite{Agashe:2014kda},
but also offer a precious opportunity to test the
Mikheyev-Smirnov-Wolfenstein (MSW) matter
effect~\cite{Mikheev:1986gs,Wolfenstein:1977ue} and to probe the
solar properties, such as the core metallicity (by measuring the
CNO neutrino flux) and the total luminosity (through determining
the $pp$ neutrino flux). Furthermore, the observations of solar
neutrinos in the future water-Cherenkov detector
Hyper-Kamiokande~\cite{Abe:2011ts} and liquid-scintillator
detectors SNO+~\cite{Kraus:2010zzb}, JUNO \cite{Li:2013zyd},
RENO50 \cite{Kim:2014rfa} and LENA \cite{Wurm:2011zn} will greatly
improve current knowledge about the electromagnetic properties of
neutrinos.

Besides reactor antineutrino experiments, the neutrino-electron
elastic scattering of low-energy solar neutrinos can also be used
to measure the neutrino magnetic properties. The contribution of
the neutrino magnetic dipole moment to the elastic $\nu_e$-$e^-$
cross section becomes more predominant as the electron kinetic
energy $T_{e}$ decreases since it is inversely proportional to
$T_{e}$ at low energy. Therefore, the measurements of solar
neutrinos in the $^8$B \cite{Liu:2004ny}, $^7$Be
\cite{Arpesella:2008mt} and $pp$ \cite{Bellini:2014uqa} processes
may provide us excellent opportunities to constrain the
neutrino magnetic dipole moment. As already reported in
Table~\ref{D055}, the current upper limits at 90\% C.L. obtained
from the measurements of $^8$B and $^7$Be neutrinos are $1.1\times
10^{-10} \mu_{\rm B}$ \cite{Liu:2004ny} and $5.4\times 10^{-11}
\mu_{\rm B}$ \cite{Arpesella:2008mt}, respectively. Note that
these are limits on effective magnetic moments which are different
combinations of the magnetic dipole moments of massive neutrinos,
as discussed at the beginning of Section~\ref{S004}.

In future, large liquid-scintillator detectors will improve the
precision of low-energy solar neutrino measurements, and can give
better limits on the magnetic dipole moment. There will be a
liquid-scintillator detector with a 20 kiloton target mass and a
high energy resolution of $3\%/\sqrt{E/{\rm MeV}}$ at JUNO, and
the LENA detector will be 2.5 times larger. In consequence, JUNO
\cite{Li:2013zyd} (LENA \cite{Wurm:2011zn}) will register about
four thousand (ten thousand) $^7$Be elastic $\nu_e$-$e^-$ events
per day in its detectable window above 250 keV, which means that
the statistical uncertainties can be negligible after years of
data-taking. Therefore, the achievable limit on the neutrino
magnetic dipole moment mainly depends on the systematics, and in
particular on the radioactive and cosmogenic backgrounds.

Another interesting solar neutrino process due to neutrino
magnetic properties is the spin-flavor precession mechanism
\cite{Giunti:2014ixa}. As discussed in Section~3, besides the
standard MSW resonant transition, there might be interesting
transitions between the left-handed and right-handed components of
solar neutrinos in the presence of the solar magnetic field. In
the case of Dirac neutrinos, the additional transition happens
between the active and sterile neutrino states and can be a
sub-leading effect in neutrino oscillation probabilities. More
interestingly, in the case of Majorana neutrinos, right-handed
neutrinos of the electron flavor produced in the spin-flavor
precession can be detected with the inverse beta decay reaction,
which can significantly reduce the singles background using the
coincidence of prompt and delayed signals of the reaction. The
recent measurement from Borexino \cite{Bellini:2010gn} constrains
the transition probability to be smaller than $1.3\times10^{-4}$
(90\% C.L.), which corresponds to an upper limit of $10^{-12}
\mu_{\rm B}-10^{-8} \mu_{\rm B}$ for the neutrino magnetic dipole
moment. Future liquid-scintillator detectors (e.g. JUNO, RENO50
and LENA) are 1-2 orders of magnitude larger than Borexino and may
improve the transition probability limits by one order of
magnitude. This observation could be free of the reactor
antineutrino background when one concentrates on the energy region
larger than 10 MeV.

{\bf Supernova neutrinos.} As is well known, the electromagnetic
dipole interaction of massive neutrinos couples left-handed
neutrinos to the right-handed ones. If neutrinos are Dirac
particles, right-handed neutrino states are sterile and can be
copiously produced in the supernova core, where large magnetic
fields may exist. While the left-handed neutrinos are trapped
inside the supernova core and come out by diffusion, the sterile
ones can freely escape from the core immediately after production.
Since the energy loss caused by right-handed neutrinos should not
shorten significantly the duration of the neutrino signal, which
has been observed by the Kamiokande-II, IMB and Baksan experiments
to be about ten seconds, one can obtain the restrictive limit on
the neutrino magnetic dipole moment $\mgm_{\nu} \lesssim 3\times
10^{-12} \mu_{\rm B}$~\cite{Raffelt:1999gv}. However, this bound
applies only to massive Dirac neutrinos, since the right-handed
states of Majorana neutrinos interact as Standard Model
antineutrinos and do not induce any extra energy loss because they
are trapped in the core.

Although it was pointed out long time ago that the
neutrino-neutrino refraction in the supernova environment may be
very important for neutrino flavor conversions, the nonlinear
evolution of neutrino flavors has recently been found to
dramatically change neutrino energy spectra~\cite{Duan:2010bg}.
Depending on the initial neutrino fluxes and energy spectra, a
complete swap between neutrino spectra of electron and
non-electron flavors can take place in the whole or a finite
energy range, as a direct consequence of collective neutrino
oscillations. The impact of nonzero transition magnetic moments
for massive Majorana neutrinos on collective neutrino oscillations
has been explored in Ref.~\cite{deGouvea:2012hg,deGouvea:2013zp}.
For a magnetic field of $10^{12}~{\rm G}$ and $\mgm_{\nu} \approx
10^{-22} \mu_{\rm B}$, which is just two orders of magnitude
larger than the Standard-Model prediction corresponding to
neutrino masses of the order of $0.1~{\rm eV}$, the pattern of
spectral splits of supernova neutrinos may be observed in future
experiments.

For a future galactic supernova, a number of large water-Cherenkov (Super-Kamiokande
\cite{Ikeda:2007sa} and Hyper-Kamiokande), scintillator (JUNO,
RENO50 and LENA) and liquid-argon (DUNE \cite{Adams:2013qkq})
detectors will be able to perform a high-statistics measurement of
galactic supernova neutrinos. In the case of a galactic supernova at a
typical distance of $10~{\rm kpc}$, the JUNO detector will record
about 5000 inverse beta-decay events, implying a precise
determination of $\overline{\nu}_e$ energy spectrum. In addition,
the charged-current interaction $\nu_e + {^{12}}{\rm C} \to e^- +
{^{12}}{\rm N}$ contributes to a few hundred events, which
together with the elastic $\nu_e$-$e^-$ scattering leads to a
possible measurement of $\nu_e$ energy spectrum. Finally, the
number of elastic neutrino-proton scattering events reaches two
thousand, since JUNO is expected to achieve a threshold around
$0.1$ or $0.2$ MeV for the proton recoil energy. Combining these measurements with the information
from the water-Cherenkov and liquid-argon
detectors, we hope to pin down the neutrino energy spectra with
reasonable accuracy.
The identification of the spectral splits
will allow us to probe values of the neutrino magnetic
moments  which are extremely small and impossible to
detect in other terrestrial experiments.
Unfortunately,
the experimental
determination of the neutrino magnetic moments
will be complicated by the distorsions
of the neutrino spectra
induced by the ordinary Mikheyev-Smirnov-Wolfenstein effects in the
supernova envelope and by the Earth matter effects.

{\bf Cosmological observations.} The early Universe is another
place where neutrinos can be in thermal equilibrium and play a
very important role. The phase transitions in the early Universe
can have generated primordial magnetic fields, which populated the
right-handed neutrinos if neutrinos are Dirac particles and have
finite magnetic dipole moments. If the magnetic dipole interaction
rate of neutrinos is larger than the Hubble expansion rate during
the epoch of primordial nucleosynthesis, the right-handed
neutrinos are in thermal equilibrium and contribute to the
effective number of neutrino species by $\Delta N^{}_\nu = 3$,
which will modify the correct predictions of the standard BBN
theory for the abundance of light nuclear elements. As shown in
Refs.~\cite{Enqvist:1994mb,Long:2015cza}, the requirement for the
magnetic dipole interaction rate to be smaller than the Hubble
expansion rate at $T = 200~{\rm MeV}$, when the QCD phase
transition occurs, leads to an upper bound on the neutrino
magnetic dipole moment. For a primordial magnetic field $B_0 =
10^{-14}~{\rm G}$ and the size of magnetic field domain $\lambda =
1~{\rm Mpc}$, one can derive a tight bound $\mgm_{\nu} <
10^{-16}~\mu_{\rm B}$, which is several orders of magnitude below
current experimental limits~\cite{Long:2015cza}.

\section{Summary and prospects}
\label{H001}

In this review we outlined some aspects of the physics of
electromagnetic neutrinos.
No experimental evidence in favor of neutrino electromagnetic
interactions has been obtained so far. All the neutrino
electromagnetic characteristics have rather stringent upper
bounds, which are due to laboratory experiments or from
astrophysical observations.

The most accessible neutrino electromagnetic property may be the
charge radius, for which the Standard Model gives a value which is
only about one order of magnitude smaller than the experimental
upper bounds. A measurement of a neutrino charge radius at the
level predicted by the Standard Model would be another spectacular
confirmation of the Standard Model, after the recent discovery of
the Higgs boson (see \cite{1312.5672}). However, such a
measurement would not give information on new physics beyond the
Standard Model unless the measured value is shown to be
incompatible with the Standard Model value in a high-precision
experiment.

The strongest current efforts to probe the physics beyond the
Standard Model by measuring neutrino electromagnetic properties is
the search for a neutrino magnetic moment effect in reactor
$\bar\nu_{e}$-$e^{-}$ scattering experiments. The current upper
bounds reviewed in Section~\ref{S004} are more than eight orders
of magnitude larger than the prediction discussed in
Section~\ref{C001} of the Dirac neutrino magnetic moments in the
minimal extension of the Standard Model with right-handed
neutrinos. Hence, a discovery of a neutrino magnetic moment effect
in reactor $\bar\nu_{e}$-$e^{-}$ scattering experiments would be a
very exciting evidence of non-minimal new physics beyond the
Standard Model.

In particular, the GEMMA-II collaboration expects to reach around
the year 2017 a sensitivity to $\mgm_{\nu_{e}} \approx 1 \times
10^{-11} \bmag$ in a new series of measurements at the Kalinin
Nuclear Power Plant with a doubled neutrino flux obtained by
reducing the distance between the reactor and the detector from
13.9 m to 10 m and by lowering the energy threshold from 2.8 keV
to 1.5 keV \cite{Beda:2012zz,Beda:2013mta}. The corresponding
sensitivity to the neutrino electric millicharge will reach the
level of $|\chg_{\nu_{e}}| \approx 3.7 \times 10^{-13} \, \elechg$
\cite{Studenikin:2013my}.

There is also a GEMMA-III project\footnote{Victor Brudanin and
Vyacheslav Egorov, private communication.} to further lower the
energy threshold to about 350 eV, which may allow the experimental
collaboration to reach a sensitivity of $\mgm_{\nu_{e}} \approx 9
\times 10^{-12} \bmag$. The corresponding sensitivity to neutrino
millicharge will be $|\chg_{\nu_{e}}| \approx 1.8 \times 10^{-13}
\, \elechg$ \cite{Studenikin:2013my}.

An interesting possibility for exploring very small values of
$\mgm_{\nu_{e}}$ in $\bar\nu_{e}$-$e^{-}$ scattering experiments
has been proposed in Ref.~\cite{Bernabeu:2004ay} on the basis of the
observation \cite{Segura:1993tu} that ``dynamical zeros'' induced
by a destructive interference between the left-handed and
right-handed chiral couplings of the electron in the charged and
neutral current amplitudes appear in the Standard Model
contribution to the scattering cross section. It may be possible
to enhance the sensitivity of an experiment to $\mgm_{\nu_{e}}$ by
selecting recoil electrons contained in a forward narrow cone
corresponding to a dynamical zero (see Eq.~(\ref{D036})).

In the future, experimental searches of neutrino electromagnetic
properties may be performed also with new neutrino sources, as a
tritium source \cite{McLaughlin:2003yg}, a low-energy beta-beam
\cite{McLaughlin:2003yg,deGouvea:2006cb}, a stopped-pion neutrino
source \cite{Scholberg:2005qs}, or a neutrino factory
\cite{deGouvea:2006cb}. Recently the authors of Ref.~\cite{Coloma:2014hka} proposed to
improve the existing limit on the electron neutrino magnetic
moment with a megacurie $^{51}\text{Cr}$ neutrino source and a
large liquid Xe detector.

Neutrino electromagnetic interactions could have important effects
in astrophysical environments and in the evolution of the Universe
and the current rapid advances of astrophysical and cosmological
observations may lead soon to the very exciting discovery of
nonstandard neutrino electromagnetic properties. In particular,
future high-precision observations of supernova neutrino fluxes
may reveal the effects of collective spin-flavor oscillations due
to Majorana transition magnetic moments as small as $10^{-21} \,
\bmag$ \cite{deGouvea:2012hg,deGouvea:2013zp}.

Let us finally emphasize the importance to pursue the experimental
and theoretical studies of electromagnetic neutrinos, which could
open a door to new physics beyond the Standard Model.

\vskip 12pt

\noindent {\small {\bf Acknowledgements.}
This work was supported in part by the joint project of the
Russian Foundation for Basic Research (RFBR) under Grant No.
15-52-53112 and National Natural Science Foundation of China
(NSFC) under Grant No. 11511130016. The work of C. Giunti was
partially supported by the PRIN 2012 research grant 2012CPPYP7.
The work of K. A. Kouzakov, A. V. Lokhov, and A. I. Studenikin was
also supported in part by the RFBR under Grant
No.~14-22-03043-ofi-m, and that of Yu-Feng Li and Shun Zhou by the
NSFC under Grant Nos. 11135009, 11305193, by the Innovation
Program of the Institute of High Energy Physics under Grant No.
Y4515570U1, and by the CAS Center for Excellence in Particle
Physics (CCEPP). K. A. Kouzakov also acknowledges support from the
RFBR under Grant No. 14-01-00420-a.}

\bibliographystyle{andp2012}
\bibliography{database2,emp}

\end{document}